\documentclass{aastex}
\usepackage{spr-astr-addons}
\usepackage{url}

\RequirePackage{color}
\usepackage{latexsym}
\usepackage{graphics}
\usepackage{graphicx}
\usepackage{epsfig}

\begin{document}

\title{ Electron capture strength on odd-A nucleus $\mathbf{^{59}}$Co in explosive astrophysical environment}
\shorttitle{Stellar capture rates on $^{59}$Co}
\author{Muneeb-Ur Rahman\altaffilmark{1}}
\affil{Department of Physics, Islamia College Peshawar, KPK,
Pakistan\\ email: muneeb@icp.edu.pk}
\author{Jameel-Un Nabi\altaffilmark{2}}
\affil{Faculty of Engineering Sciences, GIK Institute of Engineering
Sciences and Technology, Topi 23640, Swabi, KPK, Pakistan\\ email:
jameel@giki.edu.pk }

\begin{abstract}
The Gamow-Teller (GT) transitions within massive stars play
sumptuous role in the dynamics of core collapse supernovae. GT
strength distributions and electron capture rates have been
calculated for odd-A nucleus $^{59}$Co within the proton-neutron
quasiparticles random phase approximation (pn-QRPA) formalism. The
pn-QRPA results are compared with other model calculations and (n,
p) reaction experiment carried out at TRIUMF charge-exchange
facility.  The pn-QRPA calculated a total $B(GT_{+})$ strength of
3.3 for $^{59}$Co to be compared with the shell model value of 2.5
and the 1.9 $\pm$ 0.1 in the (n, p) charge-exchange reaction.
Aufderheide et. al. \cite{Aufderheide93} extracted total strength
equaling 2.4 $\pm$ 0.3. The placement of GT centroid at 5.6 MeV by
the pn-QRPA model is in reasonable agreement with the shell model
centroid at 5.1 MeV whereas the measured GT centroid was placed at
4.4 $\pm$ 0.3 MeV in the (n, p) experiment. Fuller, Fowler and
Newman (FFN) \cite{FFN80, FFN82, FFN82b}, placed the GT centroid at
too low excitation energy of 2.0 MeV in the daughter nucleus
$^{59}$Fe, and this misplacement led to the enhancement of FFN
rates. The suppressed pn-QRPA and shell model electron capture rates
are in good agreement with each other. The rates are suggestive of
higher value of $Y_{e}$ (electron-to-baryon ratio) and may
contribute to a more massive homologously collapsing core resulting
in a more energetic shock. It might be interesting for the
simulators to check the effect of these suppressed rates on the
fine-tuning of the time rate of $Y_{e}$, the concomitant heavy
element nucleosynthesis, and, on the energetics of the subsequent
shock wave.
\end{abstract}
\keywords{Gamow-Teller (GT) strength distribution; GT centroid;
electron capture; pn-QRPA; stellar dynamics; core-collapse.}

\section{Introduction}
Supernovae (either type-Ia or type-II) are very crucial to our very
existence as well as the structural and morphological development of
the galaxies and universe at large. These two types of supernovae
are the major contributors for the elements production in the
universe. Iron in the universe is made approximately in equal amount
by type-Ia and type-II supernovae \cite{Cow04}. The incarnation of
these supernovae starts when the mass of the iron core exceeds the
appropriate Chandrasekhar mass limit. The electron degeneracy
pressure is no longer in a position to sustain the inner core
against the gravitational collapse. Consequently the implosion
ensues and leads to a more exotic and denser matter. A smaller
entropy as well as a smaller iron core mass favor the explosion
mechanism as the shock has to plough a little mass of the inner
core. The shock wave looses less energy in the photodisintegration
process and the lower entropy helps in the reduction of free protons
in stellar matter which in turn reduces the probability of electron
capture rates on free protons and thus leads to a higher value of
$Y_{e}$ (lepton-to-baryon ratio) at the bounce.

The mass of the homologous collapsing core and subsequent shock is
determined by final value of the $Y_{e}$ of the stellar core. The
mass of the homologous collapsing core is related to the final
lepton number as $M_{hc} \propto Y_{ef}^2$. When the core stiffens
beyond the nuclear density it sends an outward shock wave with
energy:

\begin{eqnarray}
E_{S}\simeq (GM_{HC}^2/R_{HC})(Y_{ef}- Y_{ei})\\
\nonumber\simeq M_{HC}^{5/3}(Y_{ef}-Y_{ei}) \simeq
Y_{ef}^{10/3}(Y_{ef}-Y_{ei}),
\end{eqnarray}
disintegrating the stellar matter \cite{ Kar94}. Where $M_{HC}$,
$R_{HC}$ are the mass and radius of the homologous core,
respectively. The energy of the shock is larger for larger
difference of initial and final lepton fractions and for larger
final lepton fraction prior to collapse. Weak decay rates, such as
electron/positron captures and $\beta^{\pm}$ decays, affect the
central lepton fraction $Y_{e}$, which subsequently determines the
composition of the ejecta from supernova explosions
\cite{Brachwitz00,Nabi12,Nabi and Rahman11}. Iron group nuclei have
highest binding energy and element synthesis beyond the iron group
nuclei in the stellar kiln is energetically not possible.
Consequently the star's energy budget and electron degeneracy
pressure are unable to counter the mammoth inward gravitational
force. The density and temperature of the stellar core increase with
further addition of mass and this leads the core to reduce its free
energy via electron capture process on protons in the nuclei and
reduces the $Y_{e}$ at initial stage of the collapse
\cite{Langanke13}. The concomitant nucleosynthesis via electron
capture change the $Y_{e}$ and consequently the composition of
matter in astrophysical scenario. At this stage, when the density is
not high enough ($ \le 10^{10}$ g/cm${^3}$), the neutrinos bleed out
from the star carrying along enormous amount of energy. Both
neutrino bleeding and reduction in $Y_{e}$ accelerate the collapse
process. This collapse of the star is very sensitive to the entropy
and to the number of leptons per baryon, $Y_{e}$ \cite{Nabi and
Rahman11, Bethe79, Nabi and Rahman05, Rahman and Nabi13}.

Bethe and collaborators \cite{Bethe79} investigated, for the first
time, the importance of Gamow-Teller (GT) transitions for the
stellar weak decay rates. The GT transitions are one of the most
essential and paramount nuclear weak processes of the spin-isospin
($\sigma\tau$) type and are involved actively in various processes
such as nucleosynthesis and stellar core collapse of massive stars
preceding supernova explosions. Later Fuller, Fowler, and Newman
\cite{FFN80, FFN82, FFN82b} (referred to as FFN in this paper)
investigated the GT transitions and calculated the weak decay rates
for 226 nuclei in the mass range 21 $\le A \le $ 60 at densities and
temperatures pertinent to astrophysical environment using
parameterization based on independent particle model. For unmeasured
GT transition of allowed nature, they assumed an empirical value of
log\textit{ft} = 5. FFN employed the so-called Brink's hypothesis in
their calculations. This hypothesis assumes that the GT strength
distributions on the excited states are the same as that on the
ground state shifted only by the excitation energy of that
particular excited state (see Ref. \cite{Aufderheide91} for further
details). Brink's hypothesis was later found to be a poor
approximation to be employed in stellar weak interaction rate
calculations \cite{Nabi and Rahman11, Nabi and Rahman05,Rahman and
Nabi13, Nabi and Rahman07,Nabi09,Nabi12a}. Charge-exchange
reactions, such as (p, n) and (n, p) experiments, confirm the
misplacement of GT centroid in FFN's calculations and it was further
observed that GT strength is quenched and fragmented. These
developments led the scientific community to use microscopic
approaches for a reliable calculation of stellar weak decay rates.

Large-scale shell model \cite{Langanke00} and proton-neutron
quasiparticle random phase approximation (pn-QRPA) theory
\cite{Nabi99} are two such microscopic models used with relative
success for a reliable calculation of stellar weak decay rates. Both
microscopic models have associated pros and cons. In shell model
emphasis is more on interactions as compared to correlations whereas
pn-QRPA puts more weight in correlations. It is difficult to comment
which method should suit the notoriously complex dynamics of core
collapse. The shell model Monte Carlo (SMMC) techniques (e.g. Ref.
\cite{Dean98}) were also used for the calculations of the GT
strength distributions in \textit{pf}-shell nuclei with associated
pros and cons. SMMC does not allow for detailed nuclear spectroscopy
and has some restrictions in its applications for odd-A and odd-odd
nuclei. It has also been noted that the SMMC's average GT strength
distribution introduces inaccuracies in the weak decay rates
calculations \cite{Caurier99}. Different calculations of GT strength
distributions and weak decay rates are now available, and can be
found in literature (e.g. \cite{Nabi and Rahman11,Nabi and Rahman07,
Nabi99, Caurier99, Zhi11, Sasano12, Heger01, Fujita05}).

Electron capture rate on $^{59}$Co is argued to play a pivotal role
in the presupernova evolution of massive stars (e.g. see the
simulation results of Refs. \cite{Heger01,Auf94}. In this paper we
used the pn-QRPA theory to calculate the GT strength distributions
and weak decay rates for the odd mass nucleus, $^{59}$Co, at
densities and temperature relevant for pre-collapse and supernova
phases of massive stars.

The paper is organized as follows. Section 2 describes the
theoretical formalism used for the calculation of $B(GT_{+})$
strength and weak decay rates based on pn-QRPA model. The calculated
GT strength for $^{59}$Co is presented in Section 3. Here we also
present comparisons with the (n, p) experiment and previous
calculations. The calculated electron capture rates are presented in
Section 4. We finally conclude our findings in Section 5.

\section{Formalism}
The electron capture rates (Z, A) + e $\rightarrow$ (Z-1, A) + $\nu$
mediated by charged weak interaction are calculated within the
domain of pn-QRPA model. The pn-QRPA is considered to be an
efficient model to extract the GT strength distributions for ground
as well as excited states of the nuclei present in stellar matter
\cite{Nabi99, Nabi13}. During pre-collapse and supernova phases of
massive stars, transitions from excited states contribute
effectively to the total electron capture rate and a microscopic
calculation of \textit{all} excited state GT strength distributions
is desirable. The pn-QRPA model is used in the present work to
calculate the $B(GT_{+})$ strength distribution and associated
electron capture rates on odd-A nucleus $^{59}$Co using a luxurious
model space of $7\hbar\omega$. The model is capable of performing a
microscopic calculation of ground \textit{and} excited states GT
strength distribution functions.

The Hamiltonian used in our calculation had four parts and is of the
form
\begin{equation} \label{GrindEQ__2_}
{\rm H}^{{\rm QRPA}} {\rm \; =\; H}^{{\rm sp}} {\rm \; +\; V}^{{\rm
pair}} {\rm \; +\; V}_{{\rm GT}}^{{\rm ph}} {\rm \; +\; V}_{{\rm
GT}}^{{\rm pp}},
\end{equation}
where $H^{sp} $, $V^{pair} $, $V_{GT}^{ph} $, $V_{GT}^{pp} $ are the
single-particle Hamiltonian, the pairing force, the particle-hole
(ph) GT force, and the particle-particle (pp) GT force,
respectively. In the present work, in addition to the well known
particle-hole GT force \cite{Halbleib67,Staudt90,Muto92}, the
particle-particle GT force \cite{Soloviev87, Kuzmin88} is also taken
into account.
The electron capture rate of a transition from the
\textit{i}th state of a parent nucleus (Z, N) to the $jth$ state of
the daughter nucleus $(Z - 1, N + 1)$ is given by

\begin{equation} \label{GrindEQ__3_}
\lambda _{ij}^{ec} \, =\, \ln 2\frac{f_{ij} (T,\rho ,E_{f}
)}{(ft)_{ij} }.
\end{equation}

The $(ft)_{ij} $ of an ordinary $\beta ^{\pm } $ decay from the
state ${\left| i \right\rangle} $ of the parent nucleus to the state
${\left| f \right\rangle} $ of the daughter is related to the
reduced transition probability $B_{ij} $ of the nuclear transition
by

\begin{equation} \label{GrindEQ__4_}
(ft)_{ij} \, =\, {D\mathord{\left/ {\vphantom {D B_{ij} .}} \right.
\kern-\nulldelimiterspace} B_{ij}.}
\end{equation}

The value of D is taken to be 6146 $\pm$ 6 s adopted from Ref.
\cite{Jokinen02}, and B$_{ij}$ is given by
\begin{equation} \label{GrindEQ__5_}
B_{ij} \, =\, B(F)_{ij} \, +\, \left({g_{A} \mathord{\left/
{\vphantom {g_{A}  g_{V} }} \right. \kern-\nulldelimiterspace} g_{V}
} \right)_{\emph{eff}}^{2} B(GT)_{ij}.
\end{equation}
Here $B(F)_{ij}$ and $B(GT)_{ij}$ are the reduced transition
probabilities due to Fermi and GT transitions and $\left({g_{A}
\mathord{\left/ {\vphantom {g_{A} g_{V} }} \right.
\kern-\nulldelimiterspace} g_{V} } \right)_{\emph{eff}}^{2}$ is the
effective ratio of the axial-vector $(g_{A} )$ to the vector $(g_{V}
)$ coupling constants that takes into account the observed quenching
of the GT strength \cite{Osterfeld92}. In the present work
$\left({g_{A} \mathord{\left/ {\vphantom {g_{A} g_{V} }} \right.
\kern-\nulldelimiterspace} g_{V} } \right)_{\emph{eff}}$ was taken
as
\begin{equation} \label{GrindEQ__6_}
\left({g_{A} \mathord{\left/ {\vphantom {g_{A}  g_{V} }} \right.
\kern-\nulldelimiterspace} g_{V} } \right)_{\emph{eff}}^{2} \, =\,
0.66 \left({g_{A} \mathord{\left/ {\vphantom {g_{A}  g_{V} }}
\right. \kern-\nulldelimiterspace} g_{V} } \right)_{\emph{bare}}^{2}
\end{equation}
with $\left({g_{A} \mathord{\left/ {\vphantom {g_{A}  g_{V} }}
\right. \kern-\nulldelimiterspace} g_{V} } \right)_{\emph{bare}}$
taken as -1.257.

The phase space integrals are
\begin{equation} \label{GrindEQ__7_}
f_{ij} \, =\, \int _{w_{l} }^{\infty }w\sqrt{w^{2} -1}  (w_{m} \,
+\, w)^{2} F(+ Z,w)G_{-} dw,
\end{equation}
where $G_{-} $ is the electron distribution function. $F(+z,w)$ are
the so-called Fermi functions and are calculated according to the
procedure adopted by Ref. \cite{Gove71}. In Eq.~\ref{GrindEQ__7_}
energies are given in units of $m_{e}c^{2}$ and momenta in units of
$m_{e}c$. In Eq.~\ref{GrindEQ__7_}, $w$ is the total energy of the
electron including its rest mass, $w_{l} $ is the total capture
threshold energy (rest + kinetic) for electron capture.  One should
note that if the corresponding positron emission total energy,
$w_{m}$, is greater than -1, then $w_{l}=1$, and if it is less than
or equal to 1, then $w_{l}=\mid w_{m} \mid$. $w_{m}$ is the total
$\beta$-decay energy in units of $m_{e}c^{2}$,
\begin{equation}
 w_{m} = \frac {1}{m_{e}c^{2}}(m_{p}-m_{d}+E_{i}-E_{j}),
\end{equation}
where $m_{p}$ and $E_{i}$ are mass and excitation energies of
the parent nucleus, and $m_{d}$ and $E_{j}$ of the daughter nucleus,
respectively.

The number density of electrons associated with protons and nuclei
is $\rho Y_{e} N_{A} ,$ where $\rho $ is the baryon density, $Y_{e}
$ is the ratio of electron number to the baryon number, and $N_{A} $
is the Avogadro's number.
\begin{equation} \label{GrindEQ__8_}
\rho Y_{e} \, =\, \frac{1}{\pi ^{2} N_{A} } (\frac{m_{e} c}{\hbar }
)^{3} \int _{0}^{\infty }(G_{-}  -G_{+} )p^{2} dp,
\end{equation}
where $p\, =\, (w^{2} -1)^{1/2} $ is the electron or positron
momentum in units of $m_{e}c$ and $G_{+}$ is the positron
distribution function.  The neutrino blocking of the phase space is
not taken into account. We assume that neutrinos and antineutrinos
escape freely from the interior of the star under prevailing
physical conditions. The total electron capture rate per unit time
for a nucleus in thermal equilibrium at temperature $T$ is finally
achieved by performing a summation over all partial rates and is
given by
\begin{equation} \label{GrindEQ__9_}
\lambda _{ec} \, =\, \sum _{ij}P_{i} \lambda _{ij}^{ec}.
\end{equation}
Here $P_{i} $ is the probability of occupation of parent excited
states and follows the normal Boltzmann distribution. The summation
in Eq.~9 is carried out over all initial and final states until
satisfactory convergence in the rate calculations is achieved.

\section{Comparison of Gamow-Teller Strength Distributions for $^{59}$Co}
The \textit{fp}-shell nucleus $^{59}$Co, akin to $^{51}$V, is
considered to play a significant role in the late stages just prior
to the presupernova collapse of the stellar core in massive stars
\cite{Heger01,Auf94,Baumer03}. They used the iron-group nuclei,
including $^{44-52}$Ti, $^{47-54}$V and $^{54-64}$Co, $^{56-66}$Ni,
to calculate abundances in nuclear equilibrium and stressed that
this group of nuclei should be adequate to represent the electron
capture and beta decay processes for stellar core having $Y_{e}
\gtrsim$ 0.43. The cross section for GT transitions is directly
proportional to the cross section of electron capture on nuclei
\cite{Jackson88} and is considered as important essence in the
simulation of core collapse of heavy mass stars \cite{FFN80, FFN82,
Aufderheide91}. In this section we report the pn-QRPA calculation
for the $B(GT_{+})$ strength distribution of $^{59}$Co. The
experimental data extracted in charge-exchange reactions  show that
the GT strength is strongly quenched as compared to the independent
model value and is fragmented over many states in the daughter
nucleus (e.g. \cite{Baumer03,William95}). For the case of $^{59}$Co,
the pn-QRPA GT strengths were renormalized with a quenching factor
of 0.66 (see Eq.~6).  The pn-QRPA calculated $B(GT_{+})$ results and
it's comparison with the observed GT strength function measured in
the (n, p) reaction carried out using TRIUMF charge-exchange
facility \cite{Alford93} is shown in Fig.~1. On the other hand
Fig.~2 displays a mutual comparison between the pn-QRPA and shell
model GT strength calculation \cite{Caurier99}. In both figures the
vertical axis shows $B(GT_{+})$ strength in arbitrary units whereas
the horizontal axis represents the excitation energy in daughter
nucleus, $^{59}$Fe, in units of $MeV$. The authors in
\cite{Alford93} also carried out a shell model calculation in
truncated model spaces and employed a renormalization factor of 0.24
to compare the theoretical strength with the experimental (n, p)
data. The shell model rates, performed in one major shell, in Ref.
\cite{Caurier99} also incorporated a quenching factor of 0.74 in
their calculation.

The over all morphology of pn-QRPA calculated GT strength
distribution is satisfactory except in the high energy region when
compared with the measured strength. Both the pn-QRPA and shell
model predict very little strength above 8 MeV as compared to
measured strength. The pn-QRPA extracted a total strength of 3.32
for $^{59}$Co. The total strength observed in the (n, p) reaction is
1.9 $\pm$ 0.1 units for $^{59}$Co \cite{Alford93} whereas shell
model  calculated a total value of 2.5 \cite{Caurier99}. The authors
in Ref. \cite{Aufderheide93} quoted a total GT strength of 2.4 $\pm$
0.3 for $^{59}$Co. The pn-QRPA placed the GT centroid at 5.6 MeV,
which is in reasonable agreement with the shell model centroid
placement at 5.05 MeV \cite{Caurier99}. FFN \cite{FFN80, FFN82,
FFN82b} assumed that almost all the GT strength is concentrated in a
collective state also referred to as GT resonance (GTR). FFN  lodged
the GT resonance at too low excitation energy of 2.00 MeV in
daughter $^{59}$Fe. For further details of misplacement of GT
centroid in FFN calculation we refer to \cite{Caurier99}.  The
placement of GTR by FFN is shown by an arrow in Fig.~1 and Fig.~2.
The measured GT centroid in the charge-exchange (n, p) reaction
\cite{Alford93} resides at an excitation energy of 4.4 $\pm$ 0.3 MeV
in the daughter $^{59}$Fe. The pn-QRPA calculated GT centroids and
integrated GT strengths for odd A nuclei, $^{59}$Co and $^{51}$V,
along the electron capture direction, are compared with other model
calculations and experiments in Table~1.

\section{Electron Capture Rates on $^{59}$Co}
The stellar electron capture rates on $^{59}$Co were calculated
within the pn-QRPA formalism for temperature and density grid
relevant to pre-collapse and supernova phases of massive stars. The
mass compilation of Audi and collaborators \cite{Audi03a, Audi03b}
was used to calculate Q-value of the reaction. The reported pn-QRPA
rates for $^{59}$Co are compared with large scale shell model rates
\cite{Langanke00} and those performed by FFN \cite{FFN80, FFN82,
FFN82b} in Fig.~3. Here the vertical axis represent the calculated
electron capture rates in units of $s^{-1}$. It is noted that the
capture rates are plotted on a logarithmic scale (to base 10).
Stellar temperature $T_{9}$ is shown on the horizontal axis where
$T_{9}$ represents the temperature in units of 10$^{9} K$. The
pn-QRPA model calculated a much bigger value of the total GT
strength when compared with the shell model calculation (see
Table~1). Correspondingly the shell model rates are smaller by a
factor 8 to 20 than the pn-QRPA rates in the stellar core when
encountering a low temperature and density environment. The
microscopic calculation of GT strength of low-lying states
guarantees a reliable description of the weak rates at low
temperatures and/or densities \cite{Langanke00}, where the
individual transitions play preeminent role in the capture
processes. As the temperature of the stellar core increases the
shell model rates approaches the pn-QRPA rates. For $^{59}$Co, the
pn-QRPA placed the GT centroid at an energy 0.5 MeV higher than the
shell model centroid in daughter $^{59}$Fe. This low centroid
placement by shell model effectively contributes to the enhancement
of shell model capture rates at higher temperatures. The stellar
weak rates are one of the most important nuclear physics input
parameters for simulation of core collapse. Yet the uncertainty
involved in their calculation is considerable. What the simulators
require is a reliable calculation of these rates on a detailed
temperature-density scale pertinent to the presupernova and
supernova environments. Both shell model and pn-QRPA calculations do
employ certain approximations at various stages of their calculation
and as such the total stellar electron capture rates do differ.
These small differences provide reliable alternatives to the
simulators to model a mere 1$\%$ effect (we recall that $< 1\%$ of
the gravitational binding energy of the star goes into the kinetic
energy of the ejected envelope). Calculations by Burrows and Sawyer
\cite{Burrows98} do show that correlations significantly reduce
neutrino opacity, making more energy available to be carried to the
envelope.

The pn-QRPA electron capture rates for $^{59}$Co are also compared
with the FFN rate calculations in Fig.~3. It can be seen from the
figure that the pn-QRPA rates are enhanced at $T_{9}$ = 1 for low
densities 10$^{3}$ g/cm${^3}$ and 10$^{7}$ g/cm${^3}$. At low
densities the capture rates are sensitive to the details of GT
strength distributions. FFN employ a phenomenological calculation
and did not employ any microscopic approach to calculate the GT
strength functions from the excited states. Moreover FFN used the
Brink's hypothesis (as discussed earlier) in their calculations
(also employed by large scale shell model calculation). At higher
temperatures the FFN rates get enhanced in comparison to the pn-QRPA
rates. As temperature rises the probability of occupation of the
parent excited states increases and in the FFN calculations the
parent excitation energies were not constrained (by particle
emission processes) and this resulted in an enhancement in their
capture rates at higher temperatures. FFN did not take into account
the particle emission processes from higher excited states. As a
result the parent excited states (and resulting GT transitions using
Brink's hypothesis) extended well beyond particle emission threshold
energies. These states had a finite occupation probability at high
temperatures and consequently significant contribution to total
electron capture rates as can be seen from Eq.~\ref{GrindEQ__9_}.
Another reason for enhancement in the capture rates of FFN is due to
the placement of GT centroid at too low energy of 2.00 MeV  in
daughter nucleus (for details see Refs.
\cite{Langanke00,Caurier99,Langanke98}). In high density region
($\sim 10^{11}$ g/cm${^3}$) the FFN rates are bigger by roughly an
order of magnitude for all temperature ranges for reasons already
mentioned.

\section{Discussion and Conclusions}
The aim of the present work was to calculate the $B(GT_{+})$
strength distribution and associated electron capture rates, at
densities and temperatures that are relevant to explosive stellar
environment, for the key fp-shell nucleus $^{59}$Co. The GT
transitions largely determine electron capture rates which play a
decisive role and used as a vital input parameter in the simulations
codes of core collapse of massive stars. Consequently GT strength is
an important ingredient related to the complex dynamics of
presupernova and supernova explosion. The $B(GT_{+})$ strength
distribution for odd-A $^{59}$Co nucleus was calculated using the
pn-QRPA theory with a model space of $7\hbar\omega$. The comparison
of the pn-QRPA model with experimental results of (n, p) reaction
data, and with theoretical calculations of shell model and FFN was
made. The total GT strength and the GT centroid calculated within
the domain of pn-QRPA are in reasonable agreement with the
experimental value and reported shell model values. FFN, on the
other hand, placed the GT centroid at too low excitation energy of
2.00 MeV in daughter $^{59}$Fe which resulted in artificial
enhancement of their rates. The electron capture rates in stellar
matter are sensitive to the location of GT centroid and GT strength
distribution of these low-lying states in daughter $^{59}$Fe.
Consequently the FFN rates are enhanced both at low and high density
regions in stellar matter. The FFN rates are also enhanced at higher
temperatures. The reason for this enhancement in the FFN rates was
that particle emission processes were neglected by FFN and as such
there was no cutoff in FFN calculation for parent excitation
energies. The partial capture rates from these high-lying excited
states contributed to the total rates at high temperatures. Small
changes in the binding and excitation energies can lead to
significant modifications of the predictions for the synthesis of
elements in the stellar kline. The good agreement of the low-lying
pn-QRPA calculated GT strength with experimental results may affect
the prediction and synthesis of $^{59}$Fe as well as the evolution
timescale and dynamics of the collapsing supermassive stars.

The calculated electron capture rates on $^{59}$Co in stellar matter
are in good agreement with the large scale shell model rates for all
temperature range. Only at high density of around $\rho Y_{e} =
10^{11}$ g/cm$^{3}$ do the pn-QRPA rates get suppressed by a factor
of two. The corresponding comparison with FFN rates is not as good
and the pn-QRPA electron capture rates are suppressed by roughly an
order of magnitude in high density region. Brachwitz et al.
\cite{Brachwitz00} performed model calculations for Type-Ia
supernovae using FFN electron capture rates yielding overproduction
of iron group nuclei. In order to investigate the soft X-ray
emission of the magnetars with magnetic field strengths in excess of
the quantum critical value $B_{cr} = 4.414 \times 10^{13}G$, Gao and
collaborators \cite{Gao11a,Gao11b, Gao11c,Gao12} simulated
numerically the complete process of electron capture and discussed
its importance in the interior of the magnetars. Type-Ia supernovae
are responsible for about half of the abundances of the iron-group
nuclei in the galactic evolution \cite{Cow04}. The authors in
\cite{Liu13,Juodagalvis10} improved the previous weak decay rates
evaluations by taking into account the electron screening
corrections. They found that the capture rates get suppressed by an
order of magnitude than those of FFN due to the inclusion of
electron screening effect. The current suppressed electron capture
rates suggest larger value of $Y_{e}$ in the stellar cores. The
higher the value of $Y_{e}$ the larger will be the homologous core
(see Eq.~1) which in turn affects the hydrodynamics at bounce and
energy of the shock waves \cite{Riper88}. Consequently, a prompt
shock is created farther out with less overburden of heavier and
tightly bound nuclei \cite{Bethe90}. This means that shock waves
have to spend less energy in the dissociation of smaller iron core
for outward march. Besides $Y_{e}$, the density just outside the
iron core is also an important parameter which sets the "ram
pressure" and the shock has to overcome it before generation of any
successful supernova explosion. The density in this region is larger
by as much as 50 percent and the new models find it challenging to
produce a successful explosion \cite {Heeger01}. Notoriously enough
the explosion mechanism does not depend on a single input parameter
and requires data for a group of nuclei. The nucleosynthesis of the
iron group and other nuclei strongly depend on many input
parameters, e.g., the progenitor mass, mass cut, $Y_{e}$, explosion
energy, mixing and fallback, metalicity of progenitor, energetics of
shock, entropy of the core \cite{Umeda02}. It might be interesting
for the simulators to study how the composition of the collapsing
core and ejecta would be effected using the present and previously
reported pn-QRPA  electron capture rates and to compare the
resulting simulation results with the observed abundances.

We are in a process to calculate GT$_{+}$ transitions and electron
capture rates from ground and excited states of key
\textit{pf}-shell nuclei at densities and temperatures relevant to
astrophysical scenario specially for which experimental data is
either scarce or unavailable. The GT transitions from the excited
states of the parent nuclei have finite contributions under
prevailing physical conditions. It will be interesting to study the
contribution of a microscopic calculation of GT strength
distributions from these excited states on the total weak decay
rates and its subsequent implication in simulation codes.

\textbf{Acknowledgement}: The authors would like to acknowledge the
kind suggestions of the referee(s) which led to the correction in
the formalism and overall improvement of the manuscript.

\onecolumn
%TABLES
\begin{table}
\caption{Measured and calculated centroids and total GT strengths in
electron capture direction  for odd-A nuclei. The references are:
(a) $\rightarrow$ \cite{Alford93}, (b) $\rightarrow$
\cite{Caurier99}, (c) $\rightarrow$ \cite{Baumer03}, (d)
$\rightarrow$ \cite{Alford93}, (e) $\rightarrow$ \cite{Pov01}.}
\label{ta3}
\begin{center}
\small \begin{tabular}{c|c|cc} Nucleus &
Model & E(GT$_{+}$) & $\sum B(GT_+)$   \\
& & MeV & arb. units \\
\hline $^{59}$Co & & &\\
& EXP$^{(a)}$  &  4.4 $\pm$ 0.3  & 1.9 $\pm$ 0.1 \\
& pn-QRPA & 5.60 & 3.32\\
& SM$^{(b)}$ & 5.05 & 2.50\\
& Ref. \cite{Aufderheide93} & - & 2.4 $\pm$ 0.3\\
& Ref. \cite{FFN82} & 2.00 & -\\
\hline $^{51}$V & & &\\
& EXP$^{(c)}$  &  4.1 $\pm$ 0.4  & 0.9 $\pm$ 0.1 \\
& EXP$^{(d)}$  &  -  & 1.2 $\pm$ 0.1 \\
& pn-QRPA & 4.20 & 0.79 \\
& SM$^{(e)}$ & 4.34 & -\\
& SM$^{(b)}$ & 5.18& 1.4\\
& Ref. \cite{Aufderheide93} & - & 1.5 $\pm$ 0.2\\
& Ref. \cite{FFN82} & 3.83 & -\\
\end{tabular}
\end{center}
\end{table}

%FIGURES
\begin{figure}
\begin{center}
\includegraphics[width=6in,height=5.5in]{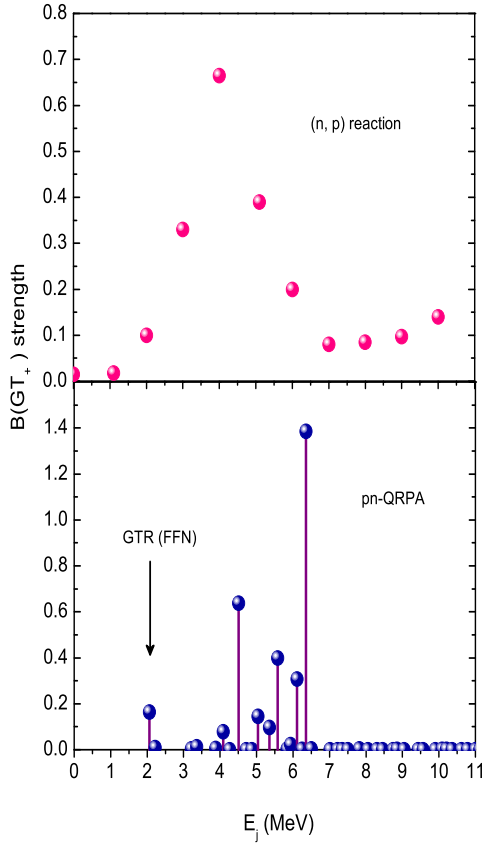}
\caption{B(GT$_{+}$) strength distributions for $^{59}$Co as
function of excitation energy in $^{59}$Fe. The upper panel depicts
result of the $(n, p)$ reaction experiment \cite{Alford93} whereas
the lower panel shows the pn-QRPA calculation (present work). The
arrow denotes the placement of the centroid of the GT resonance
predicted by FFN \cite{FFN82, FFN82b}.}\label{fig1}
\end{center}
\end{figure}
\begin{figure}
\begin{center}
\includegraphics[width=6in,height=5.5in]{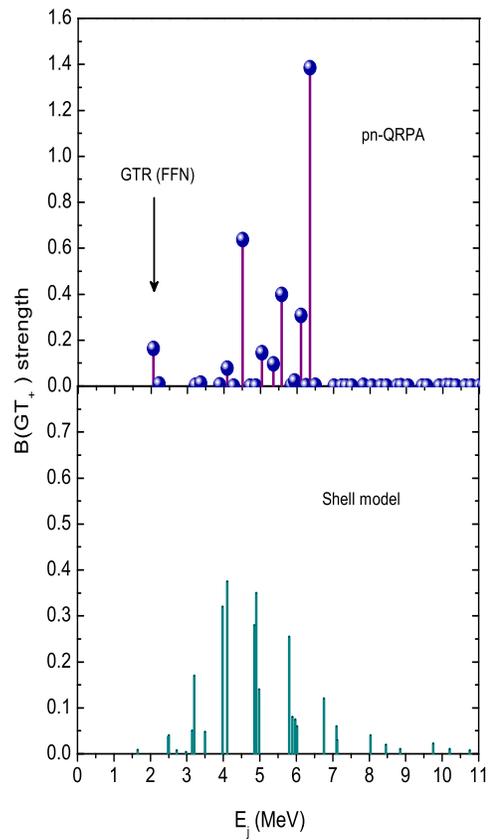}
\caption{B(GT$_{+}$) strength distributions for $^{59}$Co as
function of excitation energy in $^{59}$Fe. The upper panel depicts
the pn-QRPA calculation (present work) whereas the lower panel shows
the shell model \cite{Caurier99} results.  The arrow denotes the
placement of the centroid of the GT resonance predicted by FFN
\cite{FFN82, FFN82b}.}\label{fig2}
\end{center}
\end{figure}
\begin{figure}
\begin{center}
\includegraphics[width=6.0in,height=5.5in]{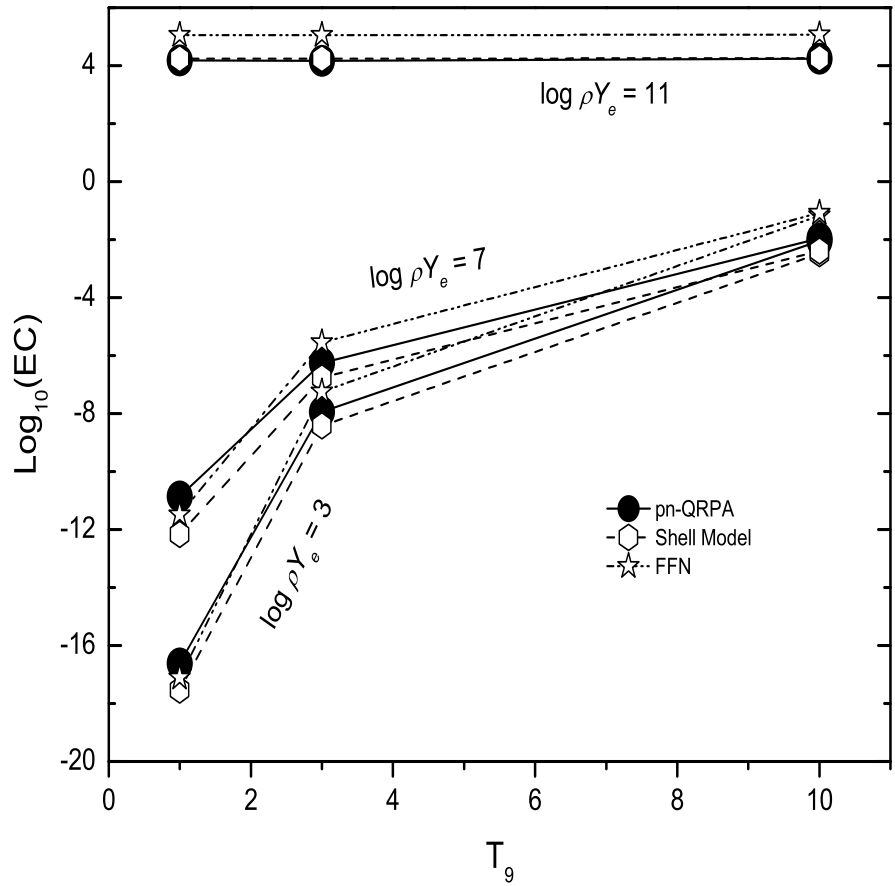} \caption { Comparison of the
pn-QRPA, large scale shell model \cite{Langanke00}, and FFN
\cite{FFN82, FFN82b} stellar electron capture rate calculations on
$^{59}$Co as a function of stellar temperature for selected
densities. For units see text.}\label{fig3}
\end{center}
\end{figure}
%\begin{figure}
%\begin{center}
%\includegraphics[width=6in,height=5.5in]{fig4.eps} \caption {Comparison of the pn-QRPA,
%and  FFN \cite{FFN82, FFN82b} stellar electron capture rate
%calculations on $^{59}$Co as a function of stellar temperature for
%selected densities. For units see text.}\label{fig4}
%\end{center}
%\end{figure}

\end{document}